\documentclass[twocolumn,aps,showpacs]{revtex4}
\usepackage[dvipdfmx]{graphicx}

\usepackage{amsmath}
\usepackage{amssymb}
\usepackage{amsthm}
\usepackage{bm}
\usepackage{latexsym}

\usepackage{color} 

\newcommand{\op}[1]{\hat {#1}}

\newcommand\ee[0]{\mathrm{e}}

\begin{document}


\title{
Nonlocal quantum gate on quantum continuous variables with minimum resources
}

\author
{Shota Yokoyama$^{1}$, Ryuji Ukai$^{1}$, Jun-ichi Yoshikawa$^{1}$, Petr Marek$^{2}$,
Radim Filip$^{2}$, 
and Akira Furusawa$^{1}$}

\affiliation{$^{1}$Department of Applied Physics, School of Engineering,
The University of Tokyo,\\ 7-3-1 Hongo, Bunkyo-ku, Tokyo 113-8656, Japan\\
$^{2}$Department of Optics, Palack\'y University, 17. listopadu 1192/12, 772 07 Olomouc, Czech Republic
}

\begin{abstract}
We experimentally demonstrate, with an all-optical setup, a nonlocal deterministic quantum non-demolition interaction gate applicable to quantum states at nodes separated by a physical distance and connected by classical communication channels.
The gate implementation, based on entangled states shared in advance, local operations, and classical communication, runs completely in parallel fashion at both the local nodes, requiring minimum resource. 
The nondemolition character of the gate up to the local unitary squeezing is verified by the analysis using several coherent states.
A genuine quantum nature of the gate is confirmed by the capability of deterministically producing an entangled state at the output from two separable input states.
The all-optical nonlocal gate operation can be potentially incorporated into distributed quantum computing with atomic or solid state systems as a cross-processor unitary operation.
\end{abstract}
\pacs{03.67.Lx, 42.50.Ex, 42.65.-k}
\maketitle

\section{Introduction}
Quantum computer is a powerful machine, capable of solving several important problems much faster than existing computers \cite{Nielsen00,Furusawa11}.
Currently, atomic or solid state quantum systems are candidates for feasible quantum computers, while
 propagating light in optical fibers is ideal for communication between quantum processors.
Small-scale quantum information processing has already been realized with various physical systems, such as superconducting qubits \cite{DiCarlo09,Yamamoto10}, trapped ions \cite{Lanyon11}, electron spins in quantum dots \cite{Koppens06}, photonic qubits \cite{Zhou13,Yao12}, and optical modes \cite{Ukai11QND,Yoshikawa07,Su13}.
Furthermore, as for the creation of multi-mode entanglement which can be exploited as a resource in quantum computation and quantum network, ultra-large-scale entangled state with more than 10,000 entangled modes was recently reported \cite{Yokoyama13}.

\begin{figure}[!b]
\centering
\includegraphics[width=8.5cm,clip]{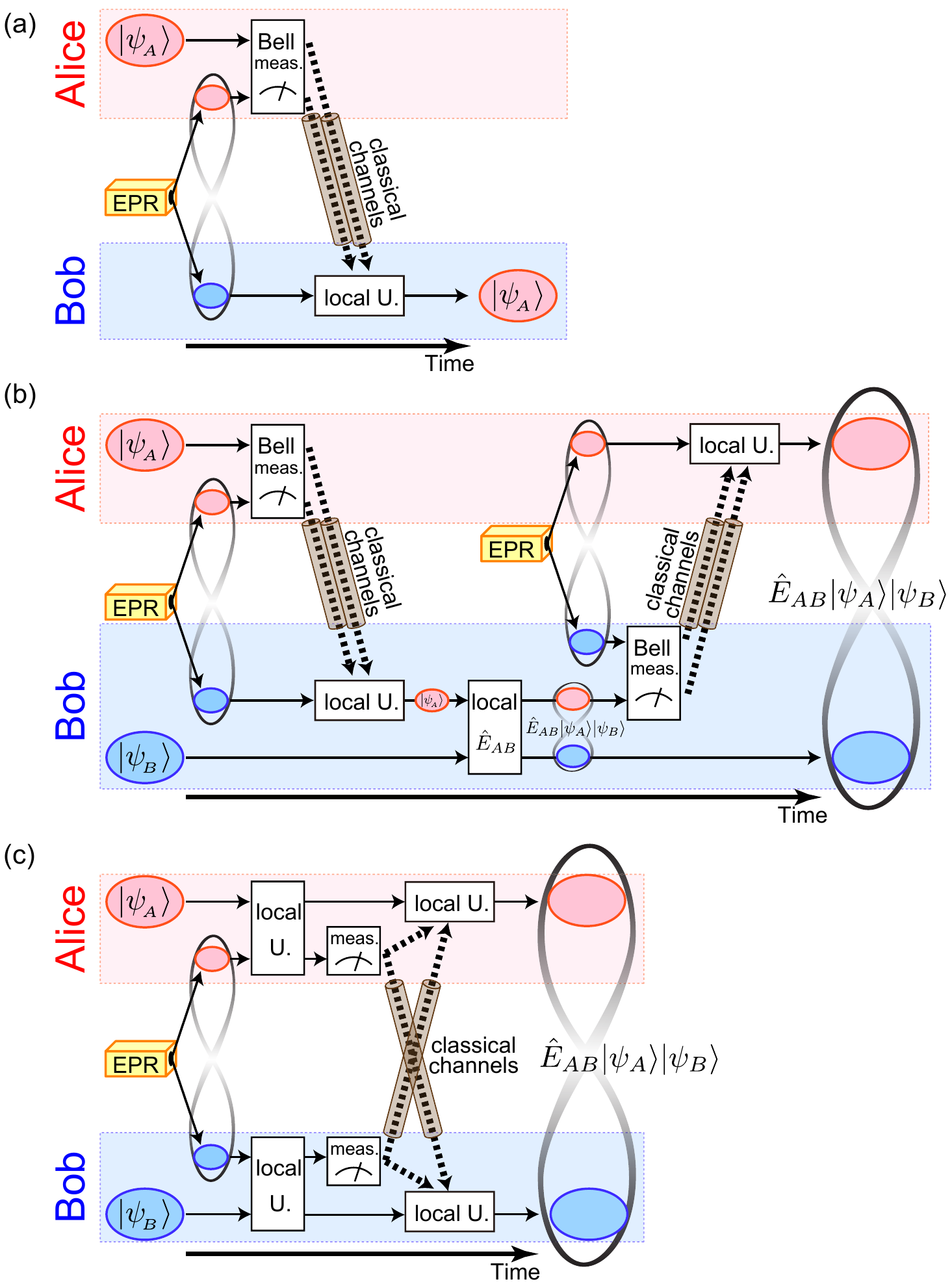}
\caption{
(color online) Abstract illustrations of some nonlocal gates.
(a) Quantum teleporatioin from Alice to Bob.
(b) Nonlocal entangling gate with sequential scheme by means of two quantum teleporters and one local entangling gate.
(c) Optimal parallel nonlocal entangling gate.
(Bell) meas., (Bell) measurement; local U., local unitary operation; EPR, entanglement resource.
}
\label{Abstract}
\end{figure}

Towards a physical implementation of a quantum computer, one approach is to make a network by connecting many quantum processors of moderate size.
If local quantum operations at the nodes are combined with quantum communication between them, any quantum processing can be decomposed to a serial combination of local operations.
In this case, the processors implement their local operations sequentially in time and use quantum channels or quantum teleporters [Fig.~\ref{Abstract}(a)] \cite{Vaidman94,Braunstein98}, to transmit quantum states among them.
 Quantum teleportation is advantageous compared to direct transfer of quantum states via quantum channels, because,
in the teleportation scenario,  the teleportation fidelity can be brought close to unity by entanglement distillation \cite{Takahashi10} even when the connecting channel is lossy.
In principle, a nonlocal quantum gate between two nodes $A$ and $B$ is achievable with a sequence of three steps [Fig.~\ref{Abstract}(b)]:  
first teleporting a state from the node $A$ to the node $B$, then locally coupling it with the other state present at the node $B$ by a local gate, and finally teleporting one outcome from that gate operation back to the node $A$.

The parallel quantum processing conceptually differs from the above-mentioned sequential one already at the level of a single basic nonlocal gate [Fig.~\ref{Abstract}(c)].
The single parallel nonlocal gate between two nodes can be built by doing all local quantum operations and classical communication (LOCC) {\it in parallel} during the same time, therefore it can be much faster than any serial method.
A reduction of operational time of nonlocal gates means less decoherence, which is the main obstacle of quantum processing.
Another advantage may be a symmetry of the parallel processing allowing for balanced use of the processors.
In practice, the resource entangled states shared in advance among nodes will be stored in quantum memories, but they can be retrieved at the time they are required and thus the subsequent implementation of the nonlocal gate itself can be deterministic within all-optical architecture, which is feasible with the current technology.

Distributed quantum computing was previously discussed for qubits \cite{Cirac99,Lo00}.
It divides tasks into subroutines and executes them in parallel at several nodes of the quantum processor network.
A parallel nonlocal gate is a nonlocal extension of the teleportation of the local gate \cite{Bartlett03}.
What are the minimal quantum resources required for implementation of a basic all-optical nonlocal gate?
For two-qubits, the basic nonlocal controlled-NOT (CNOT) gate can be principally implemented by two local CNOT gates, local projective measurements, one ebit of entanglement pre-shared between the nodes, and two-way one-bit classical communication \cite{Eisert00,Reznik02}.
The practical application of the gates for distributed quantum computing has been discussed in Ref.~\cite{Jiang07}.
Recently, the parallel implementation of the nonlocal CNOT gate has been discussed in detail \cite{Yu12}.
The heralded but still probabilistic all-optical nonlocal CNOT gates in the parallel configuration were already implemented a long time ago \cite{Gasparoni04,Huang04}.
For continuous-variable (CV) systems of quantum oscillators, an equivalent of the basic CNOT gate is a quantum non-demolition (QND) gate \cite{Filip04,Ukai11QND,Yoshikawa08}.
The non-demolition measurement based on a local QND gate is a very important topic of quantum physics \cite{Grangier98,Sewell13}.
Advantageously, QND interaction naturally appears between light and atomic memories \cite{Julsgaard04}, and therefore the QND gate is a very good candidate for a constitutive gate in CV quantum processors.
To test the principles of the basic QND gate, all optical realization of the local QND interaction for traveling beams has been constructed \cite{Yoshikawa08}.
The nonlocal realization of the QND gate in the parallel configuration and with minimal resources was actually suggested a long time ago \cite{Filip04}.
For that nonlocal QND gate, the sufficient requirements are, in analogy with the case of the qubit CNOT gate, an Einstein-Podolsky-Rosen (EPR) state pre-shared between the nodes, parallel local operations and homodyne projective measurements, and parallel two-way classical communication of a real number \cite{Filip04}.
Note that it is much less resource-consuming compared to the teleportation-based sequential strategy [Fig.~\ref{Abstract}(b)], which consumes two EPR states and requires sequentially doubled two-way classical communication.
Later, the parallel configuration of nonlocal QND gate was extended to an all-optical realization where local operations are based on local beamsplitter gates instead of local QND gates \cite{Filip05}.
In this case, the QND gate is implemented up to a priori known local squeezing operations, which were already experimentally tested on traveling optical beams \cite{Yoshikawa07} or it can be implemented directly in atomic memories \cite{Filip08}.

In this paper, we experimentally demonstrate an optimal nonlocal QND sum gate, implemented in the all-optical architecture with minimal resources according to the proposal in Ref.~\cite{Filip05}.
We demonstrate very good performance of the nonlocal QND sum gate, capable of creating entanglement represented by logarithmic negativity $E_N=0.40$ from two separable input coherent states.
The quality is limited only by the amount of shared entanglement in the EPR state (in our demonstration, it was corresponding to $-4$~dB of two-mode squeezing relative to the shot noise level).
In the ideal limit of infinite squeezing of the resource EPR state, the logarithmic negativity reaches $E_N=0.88$.
Together with the previous test of CV quantum memories \cite{Julsgaard04},  our test opens a way towards the spatially distributed parallel CV processors.

\section{Theory}
\subsection{Nonlocal QND gate with minimal resources}

We consider the scenario where two manipulators Alice and Bob, being separated by a large distance, would like to {\it simultaneously}  apply a nonlocal QND sum gate onto their states by means of LOCC, supported by pre-prepared resource entangled states.
We will consider the QND type of nonlocal deterministic sum gate described by the unitary transformation
\begin{equation}\label{qnd}
\hat \Sigma_{AB}=\ee ^{-2i\hat x_A \hat p_B},
\end{equation}
 even though the considerations can be naturally extended to other nonlocal gates.
Here $\hat x_j$ and $\hat p_j$, where $j=A,B$, denote the generalized position and momentum quadrature operators,
which satisfy the commutation relation $[\hat x_j, \hat p_k]=i \delta_{jk}/2$ with $\hbar =1/2$, where $\delta_{jk}$ is the Kronecker delta.

In Refs.~\cite{Filip04,Filip05} it was shown that one pre-shared maximally entangled state (ideal EPR state) and one ideal classical channel in each direction (two channels in total) are sufficient resources for the ideal deterministic nonlocal QND sum gate implemented in the parallel way for arbitrary input states owned by Alice and Bob.
By the classical channel we mean sending a classical real number $s\in\mathbb{R}$ over a distance. Is the maximally entangled state necessary to implement the nonlocal QND sum gate in the parallel way on the both sides for all the possible input states?
The following answer is based on the well known fact that the amount of entanglement cannot be increased by the deterministic LOCC operations \cite{Nielsen99}.
The ideal EPR states are therefore necessary to be shared between Alice and Bob, since the QND sum gate is capable of creating an entangled state with arbitrary large amount of entanglement from two pure separable states.
For instance, when the initial quantum state owned by Alice is a momentum eigenstate $|p=0\rangle$ proportional to $\int |x\rangle \mathrm{d}x$ while that owned by Bob is a position eigenstate $|x=0\rangle$, the output state after the gate operation becomes proportional to $\int |x\rangle_A\otimes|x\rangle_B \mathrm{d}x$, which is the ideal EPR state containing an infinitely large amount of entanglement and infinitely large energy.
Therefore, the ideal nonlocal QND sum gate can be only approached as the pre-shared CV entanglement between Alice and Bob infinitely increases.

To understand the role of classical communication in our procedure, we suppose that the initial state owned by Alice is a position eigenstate with an eigenvalue $x_A$ and the state owned by Bob is another position eigenstate with an eigenvalue $x_B$. Since the nonlocal gate (\ref{qnd}) implements $\hat{\Sigma}_{AB}|x_A\rangle_A \otimes |x_B\rangle_B = |x_A\rangle_A \otimes |x_B+x_A\rangle_B$, Bob could receive a classical real number $x_A$ from Alice, by comparing the position of his quantum state before and after the gate operation.
Local operations can not transmit classical information between Alice and Bob, even if they could exploit arbitrary pre-shared entanglement \cite{Peres04}.
More specifically, the maximal classical information which can be transmitted via the nonlocal quantum gate is no more than the amount of classical information required in the implementation of the nonlocal gate.
Thus Alice has to send at least one classical real number to Bob in order to implement the gate.
In a similar way, when the initial states of Alice and Bob are eigenstates of momentum,
Bob can transmit a classical real number to Alice via the nonlocal gate, which means Bob has to send at least one classical real number during the gate implementation.
Consequently, they need at least one classical channel in each direction (two channels  in total) for a nonlocal entangling gate, if it is based on LOCC and preshared entanglement.

On the other hand, if we could use high-fidelity quantum channels to directly transfer quantum states, we note that the sequential implementation requires less resources, as has been demonstrated in Refs.~\cite{Filip04,Filip05}.
In principle, it requires only a single squeezed state (while an EPR state in the parallel scheme corresponds to two squeezed states) and a single one-way classical channel.
However, the parallel scheme is still advantageous in the gate operation time.
The time cost of communication is doubled for the sequential scheme: the sequential scheme requires first quantum communication from Alice to Bob {\it after the nonlocal gate operation is required}, and second classical communication from Bob to Alice {\it after the quantum communication is completed}, while the parallel scheme requires only two-way classical communication at once during the gate operation time, because the resource EPR state can be shared {\it before the nonlocal gate operation is required}.
In addition, the implementation based on preshared resource entanglement enables entanglement distillation after they were transmitted through the quantum channels between Alice and Bob.
From the discussions above, the advantage of the parallel scheme and the minimum resource for the gate implementation is intimately connected with the possibility to preshare and distill, if quantum channels are not so reliable, quantum resources.

\begin{figure}
\centering
\includegraphics[width=8.5cm,clip]{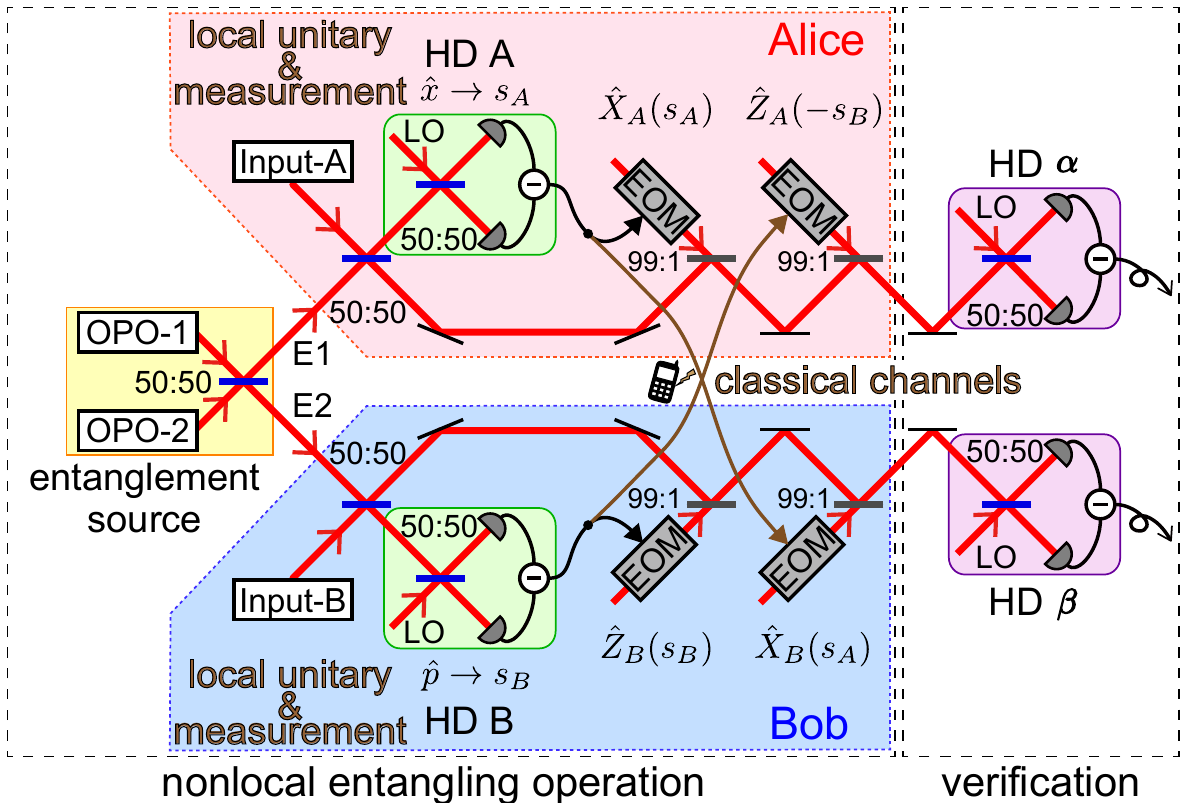}
\caption{(Color online) A schematic of our experimental setup.
OPO, optical parametric oscillator;
LO, local oscillator for homodyne measurement;
EOM, electro-optical modulator;
HD, homodyne detection;
50:50 (99:1), 50 (99)\% reflectivity beamsplitter.
}
\label{ExperimentalSetup}
\end{figure}

\subsection{Implementation of nonlocal QND gate}

The procedure of the optimal nonlocal entangling gate consists of the following three key steps \cite{Filip05}.
We will mathematically describe them by the transformations in the Heisenberg representation taking account of finite resource entanglement, which leads to a simple description for any input quantum state of the nonlocal gate.

First, the EPR state is pre-shared by Alice and Bob before the gate is actually implemented.
Note that its distribution does not reduce the speed of the gate.
To approach the ideal EPR state with both infinite energy and entanglement, we use a realistic EPR entangled state, which can be experimentally generated by combining two finitely squeezed vacuum states on a balanced beamsplitter, as is depicted in Fig.~\ref{ExperimentalSetup}. The realistic EPR state with finite entanglement is characterized by two linear combinations of position and momentum operators as below:
\begin{align}
\hat x_{E1}-\hat x_{E2} =\sqrt{2}\ee ^{-r} \hat x_1^{(0)}, \\
\hat p_{E1}+\hat p_{E2} =\sqrt{2}\ee ^{-r} \hat p_2^{(0)}.
\end{align}
Here, subscripts $E1$ and $E2$ denote two independent modes of an EPR state,
while $\ee ^{-r}\hat x_1^{(0)}$ and $\ee ^{-r}\hat p_2^{(0)}$ are squeezed quadratures of the resource modes 1 and 2 before the balanced beamsplitter combining, characterized by a squeezing parameter $r$, respectively. The limit $r\to\infty$ corresponds to  the ideal EPR state.

Second, Alice and Bob couple their own input states ($A$ and $B$, respectively) with the shared EPR state on their local balanced beamsplitters.
Then, one of the outputs on each side is measured by means of homodyne detection, making the projection on eigenstate of either position or momentum variable. The chosen measured observables correspond to
\begin{align}
\hat x_{E1'}=\dfrac{1}{\sqrt{2}} (\hat x_A -\hat x_{E1})
\quad \mathrm{and} \quad
\hat p_{E2'}=\dfrac{1}{\sqrt{2}} (\hat p_B -\hat p_{E2}) ,
\end{align}
and the measurement outcomes from them are denoted by $s_A$ and $s_B$ in the following, respectively.
On the other hand, the quadratures of the remaining parts are
\begin{align}
\hat x_{A'}=\dfrac{1}{\sqrt{2}} (\hat x_A +\hat x_{E1}),
\qquad 
\hat p_{A'}=\dfrac{1}{\sqrt{2}} (\hat p_A +\hat p_{E1}),
\\
\hat x_{B'}=\dfrac{1}{\sqrt{2}} (\hat x_B +\hat x_{E2}),
\qquad 
\hat p_{B'}=\dfrac{1}{\sqrt{2}} (\hat p_B +\hat p_{E2}).
\end{align}

Third, the measurement outcomes are transmitted to the other party through a two-way classical channel.
According to the measurement outcomes, both Alice and Bob perform feed-forward operations
expressed by the operator
$\hat X_{A'}(s_A)\hat Z_{A'}(-s_B)\hat X_{B'}(s_A)\hat Z_{B'}(s_B)$ onthe rest of their states,
where $\hat X_k(s)=\ee ^{-2is\hat p_k}$ and $\hat Z_k(s)=\ee ^{2is\hat x_k}$
are position and momentum displacement operators on modes $k=A',B'$, respectively.
Consequently, the input-output relation is given by
\begin{align}
\hat{\bm{\xi}}_{\alpha\beta}&=
\begin{pmatrix}
\sqrt{2} & 0 & 0 & 0 \\
0 & \dfrac{1}{\sqrt{2}} & 0 & -\dfrac{1}{\sqrt{2}} \\
\dfrac{1}{\sqrt{2}} & 0 & \dfrac{1}{\sqrt{2}} & 0 \\
0 & 0 & 0 & \sqrt{2}
\end{pmatrix}
\hat{\bm{\xi}}_{AB}
+\hat{\bm{\delta}}
\\
&\equiv\hat E_{AB}^\dagger \hat{\bm{\xi}}_{AB}\hat E_{AB}+\hat{\bm{\delta}},
\label{EqInOut}
\end{align}
where
$\hat{\bm{\xi}}_{AB}=(\hat x_A,\hat p_A,\hat x_B,
\hat p_B)^T$, $\hat{\bm{\xi}}_{\alpha\beta}=(\hat x_{\alpha},\hat p_{\alpha},\hat x_{\beta},
\hat p_{\beta})^T$, $\hat{\bm{\delta}}=(0,\ee ^{-r} \hat p_2^{(0)},\ee ^{-r} \hat x_1^{(0)},0)^T$
and $\hat E_{AB}$ is the entangling operator. We use indices $A,B$ for the input modes and $\alpha,\beta$ for the output modes.
In the limit of infinite squeezing $r\to\infty$, the contribution of $\hat{\bm{\delta}}$ from the pre-shared state vanishes
and the gate operation reaches a perfect unitary operation.

The obtained interaction operator $\hat E_{AB}$ can be decomposed into $\hat E_{AB} = \hat S^\dagger_A\hat S_B\hat \Sigma_{AB}$,
where $\hat S_j=\ee^{i\ln2 (\hat x_j \hat p_j +\hat p_j \hat x_j)/2}$ denotes the $-3.0$-dB $x$-squeezing operator on mode $j$.
Note that local unitary operations do not consume nonlocal resources and do not change the amount of entanglement between the two systems.
In this sense, our nonlocal gate $\hat E_{AB}$ is equivalent to a nonlocal QND sum gate $\hat \Sigma_{AB}$.
The additional local squeezing is explained from
the fact that the measurement-induced input coupling with the balanced beamsplitter
works as a universal squeezer with the squeezing level of $-3.0$~dB \cite{Filip05,Yoshikawa07,Ukai14}.
It is a difference from the implementation based on the local QND interactions \cite{Filip04}, for which the nonlocal QND is obtained without any local corrections. If it is necessary, the additional local squeezing can be corrected optically by the universal squeezers \cite{Filip05,Yoshikawa07,Ukai14}, or it can be eliminated in the quantum memory \cite{Filip08}.

Up to the residual local squeezing, the implemented nonlocal QND interaction is
\begin{equation}
\hat \Sigma_{AB}^\dagger \hat{\bm{\xi}}_{AB}\hat \Sigma_{AB}+\hat{\bm{\eta}},
\end{equation}
where $\hat{\bm{\eta}}=(0,\sqrt{2}\ee ^{-r} \hat p_2^{(0)}, \sqrt{2}\ee ^{-r} \hat x_1^{(0)},0)^T$.
The residual noise existing in $\hat p_{\alpha}$ and $\hat x_{\beta}$ variables can be reduced by sufficient squeezing from the squeezers OPO-1 and OPO-2, depicted in Fig.~\ref{ExperimentalSetup}.
Advantageously, the feed-forward corrections eliminate the noise existing in the anti-squeezed quadratures $p_1^{(0)}\ee^{r}$ and $x_2^{(0)}\ee^{r}$, therefore an impurity of the squeezed states from the OPO-1 and OPO-2 is not limiting.
The application of the high squeezing stimulates further experimental investigation of the limits of optical squeezing generated from the modern optical parametric oscillators \cite{Takeno07,Mehmet10}.

The scheme is also advantageous in the sense of efficient use of arbitrary weak resource squeezing from OPO-1 and OPO-2.
To demonstrate this, we consider the case where both input states are coherent states.
Since Eq.~\eqref{EqInOut} is linear in position and momentum operators, the output state becomes a Gaussian state.
The first moments do not affect the amount of entanglement and thus we can solely concentrate on the second central moments which are uniquely described by covariance matrix given by
$V\equiv
\tfrac{1}{2}
\bigl\langle \{
\hat{\bm{\xi}} 
,
\hat{\bm{\xi}} 
\}
\bigr\rangle$,
where $\{\hat{\bm{u}}, \hat{\bm{v}}\}\equiv \hat{\bm{u}} \hat{\bm{v}}^T +\bigl(\hat{\bm{v}} \hat{\bm{u}}^T\bigr)^T$ \cite{Menicucci11}.
Logarithmic negativity $E_N$ is a good indicator of Gaussian entanglement, invariant under local unitary operations \cite{Vidal02}.
For a Gaussian state, the logarithmic negativity can be calculated from its covariance matrix \cite{Pirandola09}.
In our case, the covariance matrix and the logarithmic negativity are as follows:
\begin{align}
V_{\alpha\beta}&=\dfrac{1}{4}
\begin{pmatrix}
2 & 0 & 1 & 0\\
0 & 1+\ee^{-2r} & 0 & -1 \\
1 & 0 & 1+\ee^{-2r} & 0 \\
0 & -1 & 0 & 2
\end{pmatrix},
\label{CMtheory}\\
E_N&=-\log \bigl(\sqrt{2(1+\ee ^{-2r})}-1\bigr)\label{LN}.
\end{align}
Note that the local squeezing unitary operations potentially used to obtain the exact QND form of the non-local interaction do not change the amount of generated entanglement.
From Eq.~\eqref{LN}, we know that $E_N$ is nonzero (positive), which means inseparability of the two subsystems even if the resource squeezing is infinitesimal.
This fact highlights efficiency of our scheme even with weak squeezing.
Note that the recent report of one-way scheme with a four-mode linear cluster state \cite{Ukai11QND} can also be regarded as a nonlocal CV gate.
However, in contrast with the current efficient and optimal scheme, that gate requires four single-mode squeezing resources with more than $-4.0$~dB in order to obtain entangled output state from two coherent inputs.

\section{Experimental setup}

The schematic of our experimental setup is shown in Fig.~\ref{ExperimentalSetup}.
The light source is a continuous-wave Ti:sapphire laser
with a wavelength of 860 nm and a power of about 1.7 W.
The quantum states to be processed are in optical modes at 1 MHz sidebands of the laser beam.

The resource EPR  beams are prepared by combining two squeezed vacuum states on a 50:50 beamsplitter.
The two squeezed vacuum states are each generated by a subthreshold optical parametric oscillator (OPO).
The OPO is a bow-tie shaped cavity with a round-trip length of 500 mm, containing a periodically poled KTiOPO$_4$ (PPKTP) crystal with 10 mm in length as a nonlinear medium \cite{Suzuki06}.
A second harmonic light-beam with the wavelength of 430 nm and the power of about 80 mW pumps each OPO, which is generated  by another bow-tie shaped cavity containing a KNbO$_3$ crystal as a nonlinear medium (omitted in Fig.~\ref{ExperimentalSetup}).
Squeezing levels of the resource squeezed vacuum states are about $-4$~dB relative to the shot noise level.

Nonzero amplitude of an input coherent state at 1 MHz sidebands is generated by modulating the optical carrier with a piezoelectric transducer (PZT) at 1 MHz.
Input coupling with  one of the EPR  beams at each party is achieved via a 50\%BS.
Then, one of the two beams in each party is measured by a  homodyne detector.
The measurement outcomes are sent to the other party, where they are used to drive suitable displacement operations.
A displacement operation is achieved by combining the main beam, which carries the quantum state to be displaced at 1 MHz, with an auxiliary beam, which is modulated at 1 MHz on a 99\%-reflectivity beamsplitter.
The auxiliary beam is modulated by an electro-optic modulator (EOM), to which the homodyne signal is sent after adjusting the gain and the phase at 1 MHz by electronic circuits.

In order to characterize the input and output states of the gate, we measure powers of the quadratures with a spectral analyzer.
The measurement frequency is 1 MHz, while the resolution and video bandwidths are 30 kHz and 300 Hz, respectively.
The data are averaged 20 times.

The propagation losses from the OPOs to the homodyne detectors are 3\% to 10\%.
The detectors' quantum efficiencies are 99\%.
The interference visibilities are 97\% on average.

\section{Experimental results}

\begin{figure*}
\centering
\includegraphics[width=16cm,clip]{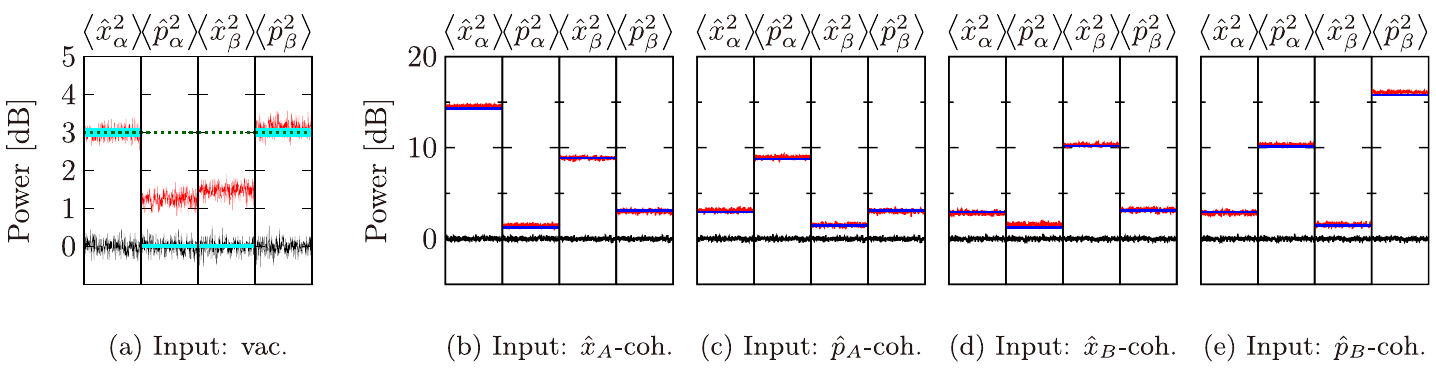}
\caption{(Color online) Powers at the outputs.
(a) For vacuum inputs.
The black and red traces show the shot noise  and experimental output quadratures, respectively.
The green dashed lines show the theoretical predictions without resource squeezing, while the cyan lines show the theoretical predictions for an ideal gate.
(b)-(e) For coherent inputs where
($\langle \hat x_A \rangle,\langle \hat p_A \rangle,\langle \hat x_B \rangle,\langle \hat p_B \rangle$)
corresponds to
$(a,0,0,0)$, $(0,a,0,0)$, $(0,0,b,0)$ and $(0,0,0,b)$, respectively.
The coherent amplitude $a$ and $b$ correspond to 11.0 dB and 12.5 dB above the shot noise level, respectively.
The blue lines show the theoretical predictions based on the experimental results of (a).
vac., vacuum state;
coh., coherent state.
}
\label{FigCoherentResults}
\end{figure*}

\begin{figure}[!t]
\centering
\includegraphics[width=7cm,clip]{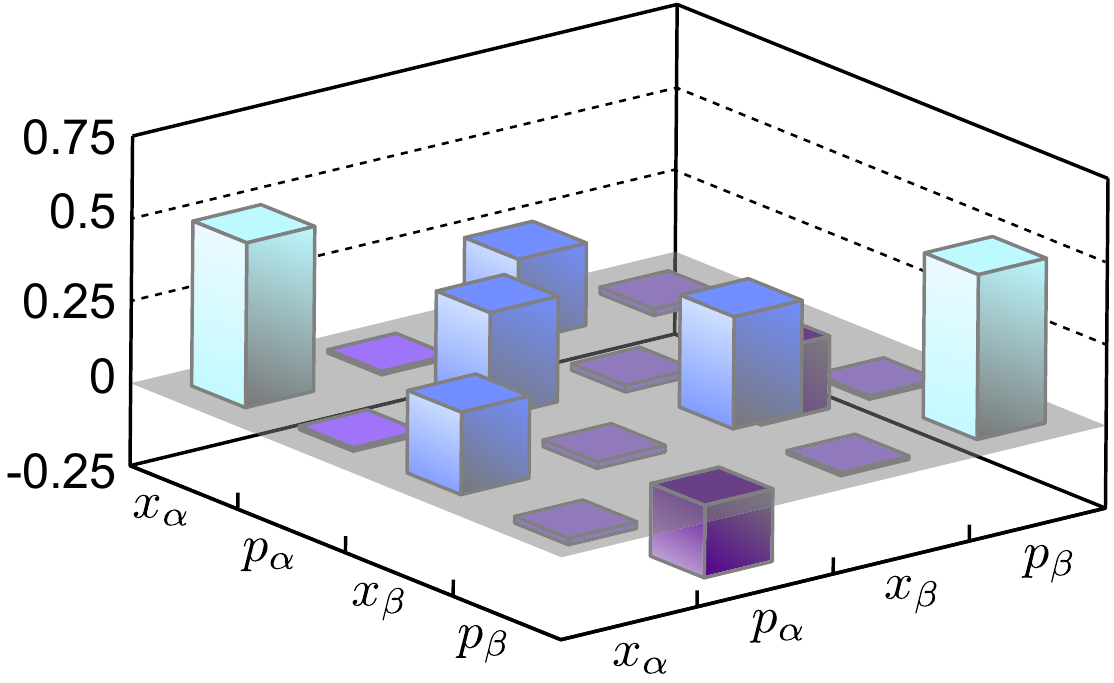}
\caption{(Color online)  A covariance matrix of output states for vacuum inputs. Measured values of the covariance matrix are shown in Eq.~\eqref{CMResult}.}
\label{FigCM}
\end{figure}

We have collected our experimental results testing the nonlocal QND sum gate up to the local squeezing operations as shown in Figs.~\ref{FigCoherentResults} and \ref{FigCM}.
Figure~\ref{FigCoherentResults} shows
the output quadrature powers for several input coherent states, from which the input-output relation is confirmed.
Figure~\ref{FigCM} shows
 a covariance matrix of one of the output Gaussian states, from which the existence of entanglement is verified.

First, we determine variances of the output quadratures by checking the case of vacuum input states and depict them in Fig.~\ref{FigCoherentResults}(a).
The theoretical predictions and experimental results are shown in Fig.~\ref{FigCoherentResults}(a).
The ideal case, which corresponds to $r\to\infty$ in Eq.~\eqref{EqInOut}, is shown by cyan lines.
On one hand, the uncorrelated quantum fluctuations of $\hat{p}_B$ and $\hat{x}_A$ are added to those of $\hat{p}_A$ and $\hat{x}_B$ by the sum gate $\hat{E}_{AB}$, which leads to 3.0~dB increase of $\hat{p}_A$ and $\hat{x}_B$.
On the other hand, additional local squeezing operation $\hat{S}_A^\dagger\hat{S}_B$ increases $\hat{x}_A$ and $\hat{p}_B$ by 3.0~dB while decreases $\hat{p}_A$ and $\hat{x}_B$.
In total, at the output of the gate, the variances of $\hat{p}_\alpha$ and
$\hat{x}_\beta$ are equal to the shot noise level (SNL) while
the variances of $\hat{x}_\alpha$ and $\hat{p}_\beta$ are 3.0 dB above the SNL (2 times the SNL).

When the resource squeezing in the OPO-1 and OPO-2 is finite, the output states are influenced by the additional excess noise.
We show as a reference the theoretical prediction for the case without the entanglement [$r = 0$ in Eq.~\eqref{EqInOut}] by green dashed lines.
In this case, the nonlocal operation is performed purely classically, assisted only by the two-way classical communication. The variances of $\hat{p}_\alpha$ and $\hat{x}_\beta$ become 3.0 dB above the SNL (2 times the SNL), while those of $\hat{x}_\alpha$ and $\hat{p}_\beta$ are not affected by the level of resource squeezing.
The experimental results of
$\langle\op{x}^2_\alpha\rangle$,
$\langle\op{p}^2_\alpha\rangle$,
$\langle\op{x}^2_\beta\rangle$, and
$\langle\op{p}^2_\beta\rangle$,
shown by the red traces, are between the cyan and green lines due to the finite resource squeezing.
They correspond to 3.0 dB, 1.2 dB, 1.5 dB, and 3.1 dB above the SNL from left to right, respectively.
These results are consistent with the resource squeezing level of $-4$~dB, which leads to 1.5 dB above the SNL for $\hat{p}_\alpha$ and $\hat{x}_\beta$.

Second, we replace the input vacuum state of either mode $A$ or mode $B$ by a coherent state, by which the input-output relation is confirmed on the assumption of linear response of the gate.
The powers of the input amplitude quadratures are individually measured in advance, corresponding to 11.0 dB for mode $A$ and 12.5 dB for mode $B$, respectively, compared to the SNL.
Figure~\ref{FigCoherentResults}(b) shows the powers of the output quadratures as red traces when the input $A$ is in a coherent state with a nonzero coherent amplitude only in the $\hat{x}_A$ quadrature.
It corresponds to 11.0 dB above the SNL, while the input $B$ is in a vacuum state.
We observe an increase in power of $\hat{x}_\alpha$ and $\hat{x}_\beta$ compared to the case of two vacuum inputs, which is caused by the nonzero coherent amplitude.
On the other hand, $\hat{p}_\alpha$ and $\hat{p}_\beta$ are not changed.
In the same figure, the theoretical prediction calculated from the measured input coherent amplitude is shown by blue lines.
Due to the additional local squeezing operators $\hat{S}_A^\dagger\hat{S}_B$, the coherent power of $\hat{x}_\alpha$ increases by 3.0 dB (corresponding to about 14 dB above the SNL), while that of $\hat{x}_\beta$ decreases by 3.0 dB (corresponding to about 8 dB above the SNL), respectively. Similarly, Figures.~\ref{FigCoherentResults}(c-e) show the results with a nonzero coherent amplitude in the $\hat{p}_A$,
$\hat{x}_B$, and $\hat{p}_B$ quadratures, respectively.

These experimental results are in good agreement with the theory described by the transformation (\ref{EqInOut}).
We see the expected feature of the sum gate that the sum of $\hat{x}_A$ and $\hat{x}_B$ appears in $\hat{x}_\beta$ while the sum of $\hat{p}_A$ and $-\hat{p}_B$ appears in $\hat{p}_\alpha$, up to the local squeezing.
We believe that the small discrepancies between our experimental results and the theoretical predictions are caused by the (unbalanced) propagation losses and nonunity visibilities of interferences with local oscillators at the homodyne detections. 

Finally, Figure~\ref{FigCM} shows the covariance matrix of the output state,
calculated from the experimental variances for the case of the vacuum input states.
The diagonal elements are obtained by measuring the variances of
the output quadratures $\hat x_j$ and $\hat p_j$.
The off-diagonal elements in each single mode, such as $V_{12}$, are obtained
by measuring the variances of $(\hat x_j\pm\hat p_j)/\sqrt{2}$.
The other off-diagonal elements can be obtained by measuring
the variances of $\hat \xi_j \pm \hat \xi_k$,
where $\hat \xi=\{\hat x,\hat p\}$.
The experimental covariance matrix was as follows
\begin{align}
V=
\left(
\begin{array}{rrrr}
0.50 & 0.01 & 0.25 & -0.02 \\
0.01 & 0.32 & -0.02 & -0.22 \\
0.25 & -0.02 & 0.34 & -0.01 \\
-0.02 & -0.22 & -0.01 & 0.50
\end{array}
\right).
\label{CMResult}
\end{align}
The margin of error for each measured matrix element is plus or minus less than 0.002.
Note that it satisfies the physical condition $V+(i/4)\ \Omega \ge 0$, where
$\Omega =\left(\begin{smallmatrix}
0 & -1 \\
1 & 0
\end{smallmatrix}
\right)
\oplus
\left(\begin{smallmatrix}
0 & -1 \\
1 & 0
\end{smallmatrix}
\right)$ \cite{Simon94,Pirandola09}.
The covariance matrix obtained from the measurement results is in good agreement with the theoretical prediction in Eq.~\eqref{CMtheory}.
We then calculate the logarithmic negativity $E_N$
of the output state from this covariance matrix by using Eq.~\eqref{CMResult}, and the obtained value is
\begin{align}
E_N&= 0.40\pm 0.01,
\end{align}
which corresponds to $-4.1 \pm 0.1$~dB of the resource squeezing.
The nonzero (positive) value is the evidence of the entanglement between the two output modes.

\section{Conclusion}
We have experimentally demonstrated an all-optical nonlocal QND sum gate for continuous variables up to local squeezing unitary operations.
Advantageously, this all-optical scheme needs only local passive beamsplitter coupling between the optical modes at each node. It also requires one pre-shared state with the EPR quantum correlations and one two-way classical channel, which are the minimal resource requirement for a nonlocal entangling QND gate.
In our experiment, all the local operations, measurement, and two-way classical communication are running truly in parallel, which increases the speed of the gate to the limit given by the technical issues. The capability of the gate to produce entanglement at the output is verified by the logarithmic negativity for the case of two coherent input states.
The nonlocal all-optical QND gate can be, together with quantum memories, in future incorporated into distributed quantum computing as a cross-processor operation.

\section*{ACKNOWLEDGMENTS}
This work was partly supported by PDIS, GIA, APSA, and FIRST initiated by CSTP,
ASCR-JSPS, and SCOPE program of the MIC of Japan.
S.Y.\ acknowledges support from ALPS\null.
R.U.\ acknowledges support from JSPS\null.
P.M.\ acknowledges support from P205/12/0577 of the Czech Science Foundation and R.F.\ support of
the Czech-Japan bilateral grant of MSMT CR No. LH13248 (KONTAKT).

\end{document}